\documentclass[superscriptaddress,nofootinbib,prd,amsmath,twocolumn]{revtex4}
\usepackage{graphicx}
\usepackage{subfigure}

\def\beq#1 #2\eeq{\begin{equation}\label{#1}#2\end{equation}}
\def\beqa#1 #2\eeqa{\begin{eqnarray}\label{#1}#2\end{eqnarray}}
\def\be#1\ee{\begin{equation}#1\end{equation}}

\newcommand\ieetm{IEEE.~Trans.~Microwave~Theory~Tech.}

\newcommand\apjs{Astrophys.~J.~Supp.}

\newcommand\apjl{ApJ~Lett.}
\newcommand\mnras{MNRAS}

\newcommand{\app}{\ensuremath{\sim}}
\def\order#1{${\cal O}\left(#1\right)$}
\newcommand{\sct}[1]{\S\ref{#1}}

\newcommand{\ka}{\ensuremath{K_a}}

\newcommand{\polar}{POLAR}
\newcommand{\polars}{POLAR's}

\newcommand\arcdeg{\mbox{$^\circ$}}%
\newcommand\arcmin{\mbox{$^\prime$}}%
\newcommand{\eqn}[1]{Equation (\ref{#1})}

\newcommand{\tbl}[1]{Table \ref{#1}}
\newcommand{\fig}[1]{Figure \ref{#1}}
\newcommand{\ie}{{\frenchspacing\it i.e.}}
\newcommand{\eg}{{\frenchspacing\it e.g. }}

\newcommand{\I}{{\bf I}}

\newcommand{\PP}{{\bf P}}
\newcommand{\M}{{\bf M}}
\newcommand{\C}{{\bf C}}

\newcommand{\D}{{\bf D}}

\newcommand{\eval}[1]{\ensuremath{\langle #1 \rangle}}

\renewcommand{\deg}{\ensuremath{^\circ}}

\newcommand{\uK}{\ensuremath{\mu K}}

\newcommand{\onephi}{\ensuremath{1\phi}}
\newcommand{\twophi}{\ensuremath{2\phi}}
\newcommand{\onephir}{\ensuremath{1\phi_r}}
\newcommand{\twophir}{\ensuremath{2\phi_r}}
\newcommand{\Zeta}{\ensuremath{\zeta}}

\newcommand{\vect}[1]{{\bf #1}}
\newcommand{\mmatrix}[1]{{\bf #1}}

\newcommand{\y}{\vect{y}}
\newcommand{\n}{\vect{n}}
\newcommand{\x}{\vect{x}}
\newcommand{\xt}{\tilde{\x}}

\newcommand{\eO}{\ensuremath{\vect{e_0}}}
\newcommand{\yeff}{\vect{y}_{pw}}
\renewcommand{\a}{\vec{a}}
\newcommand{\q}{\vect{q}}
\renewcommand{\u}{\vect{u}}

\newcommand{\A}{\mmatrix{A}}
\newcommand{\N}{\mmatrix{N}}
\newcommand{\W}{\mmatrix{W}}
\newcommand{\Z}{\mmatrix{Z}}

\newcommand{\NN}{\mmatrix{\Sigma}}

\newcommand{\Sc}{\mmatrix{S}} 
\newcommand{\Sig}{{\bf \Sigma}}
\newcommand{\Aeff}{\mmatrix{A}_{pw}}
\newcommand{\mm}{Minimum Variance Mapmaking}

\newcommand{\expec}[1]{\ensuremath{\langle #1 \rangle}}

\newcommand{\ft}{\mathcal{F}}
\newcommand{\fft}[1]{\ft \{ #1 \}}

\renewcommand{\L}{\ensuremath{{\bf \mathcal{L}}}}

\renewcommand{\l}{\ell}
\newcommand{\Cl}{C_\l}

\begin{document}

\title{CMB Polarization at Large Angular Scales: Data Analysis of the POLAR Experiment}

\author{Christopher W. O'Dell}
\affiliation{Department of Astronomy,
University of Massachusetts, Amherst, MA 01003, USA,
codell@astro.umass.edu}
\author{Brian G. Keating}
\affiliation{Division of Physics, Math, and Astronomy,
California Institute of Technology, Pasadena, CA 91125, USA}
\author{Angelica de Oliveira-Costa}
\affiliation{Department of Physics,
University of Pennsylvania, Philadelphia, PA 19104, USA}
\author{Max Tegmark}
\affiliation{Department of Physics,
University of Pennsylvania, Philadelphia, PA 19104, USA}
\author{Peter T. Timbie}
\affiliation{Department of Physics, 2531 Sterling
Hall, University of Wisconsin -- Madison, Madison, WI 53706, USA}

\begin{abstract} The coming flood of CMB polarization
experiments, spurred by the recent detection of CMB polarization
by DASI and WMAP, will be confronted by many new analysis tasks specific to
polarization.  For the analysis of CMB polarization data sets, the
devil is truly in the details. With this in mind, we present the
details of the data analysis for the POLAR experiment, which
recently led to the tightest upper limits on the polarization of
the Cosmic Microwave Background Radiation at large angular scales.
We discuss the data selection process, mapmaking algorithms,
offset removal, and the likelihood analysis which were used to
find upper limits on the polarization. Stated using
the modern convention for reporting CMB Stokes parameters,
these limits are 5.0 \uK\ on both $E$-type
and $B$-type polarization at 95\% confidence.
Finally, we discuss simulations used to test our analysis techniques
and to probe the fundamental limitations of the experiment.
\end{abstract}

\date{December 23, 2002. Submitted to Phys. Rev. D.}

\maketitle

\section{Introduction}
The detection of polarization in the Cosmic Microwave Background
(CMB) has been a long sought goal for cosmology.  CMB polarization
was recently detected at sub-degree angular scales by the DASI
instrument, a ground-based interferometer \cite{dasi02a}, and the
temperature-polarization correlation has now been detected
at larger angular scales by the WMAP satellite \cite{kogut03}. While
information in the CMB polarization at degree and sub-degree
angular scales further confirms the standard cosmological model,
polarization information at larger angular scales has the
potential to provide additional information regarding the
formation and evolution of the universe.  The process of
reionization leaves a characteristic signature on CMB polarization
at large angular scales which can be used as a means to determine
the epoch of reionization \cite{matias97,hu2000,kap02}. The power
spectrum of CMB polarization at angular scales greater than a
degree or so is sensitive to inflationary model parameters such as
the inflaton potential and the energy scale of inflation, as well
as primordial gravitational waves \cite{zs97,kks,kin98,knox02,kck03}.

These exciting rewards, taken together with increasingly sensitive
receiver technologies, have set the stage for a host of new CMB
polarization experiments.  These experiments will be faced with
new and more challenging analysis tasks than for the simpler case
of anisotropy, and it is with that in mind that we now set out to
report the details of the data analysis of the \polar\ experiment.

\polar\ (Polarization Observations of Large Angular Regions) was
designed to measure the polarization of the CMB on large angular
scales, in the \ka\ frequency band from $26-36$ GHz. This
HEMT-based correlation polarimeter operated for a single season in
the Spring of 2000 near Madison, Wisconsin. The data from this
single season led to simultaneous upper limits on E and B-type CMB
polarization, results which were initially presented in \cite{kea01}
(hereafter K01).
The details of the \polar\ instrument and its operation were later
described in \cite{kea02} (hereafter K02). In this paper we present
the details of the data selection and analysis techniques
used to arrive at the results in K01.  We also discuss the results
of a recent cross-correlation analysis of the POLAR data with
anisotropy data from the COBE-DMR experiment, described fully in
\cite{adoc02}.

The rest of this paper is organized as follows.  We first briefly
discuss the two conventions used to quantify CMB polarization.
We next review the basic properties of the POLAR instrument in \sct{s:instr}. In
\sct{s:cuts}, we discuss the data selection procedures that were
used to remove large amounts of contaminated data.
\sct{s:mapmaking} describes the mapmaking algorithms used to
transform the raw data into sky maps of $Q$ and $U$, and provides
a full pipeline simulation in order to test the algorithms.
\sct{s:likes} presents the likelihood analysis used to arrive at
the upper limits on CMB polarization, as well as the evaluation of
the polarization power spectra and some commentary on the lack of
substantial foreground contamination. Finally, in
\sct{s:discussion}, we discuss the limitations of our experiment
that could be improved upon in future projects.

\section{Q and U Conventions}\label{s:conv}
In recent years there have appeared in the literature two conventions for
reporting the Stokes parameters $Q$ and $U$ as applied to the CMB.
The first convention takes $Q(\vec{\x}) = T_x(\vec{\x}) - T_y(\vec{\x})$,
where $T_x$ and $T_y$ are antenna temperatures measured in
orthogonal directions by a
single-mode radiometer with unit optical efficiency, and $\vec{\x}$
denotes the angle of observation; $U$ is obtained similarly by
rotating the radiometer coordinates by 45\deg. These antenna
temperatures are then converted to thermodynamic temperatures in
the usual way (see \eg\ \cite{partridge95}).  This convention was
employed by the pioneering experiments of Lubin and Smoot and others
\cite{ls81,nanos79}, and was also adopted for POLAR.

The more recent convention (see \eg\ \cite{hedmanthesis,farese03}) takes $Q
= (T_x - T_y)/2$, and also has been adopted by several
experiments \cite{hed00,dasi02a,cbipc,hinshaw03}.  The results
herein are presented using the former convention (in thermodynamic
units) unless otherwise noted\footnote{We shall switch to the latter convention
when discussing the temperature-polarization correlation results.}.
Specifically, this means that our basic
limits on $E$ and $B$ type polarization (presented in K01 and in \sct{s:fbp} of this paper)
must be divided by a factor of two when directly comparing to experiments or theory
employing the latter convention. We originally reported our
simultaneous upper limits
on $E$-type and $B$-type polarization in K01 as 10.0 \uK\ at 95\% confidence;
stated using the latter convention, these limits become 5.0 \uK\ each.

\section{Instrument} \label{s:instr}
\polar\ observed the local zenith from the University of Wisconsin
-- Madison's Pine Bluff Observatory in Pine Bluff, Wisconsin
(Latitude +43\arcdeg01\arcmin, Longitude +89\arcdeg45\arcmin)
using a simple drift-scan strategy, with a 7\deg\ FWHM beam
defined by a \ka-band microwave feed horn. A correlation
radiometer operated as a polarimeter that was instantaneously
sensitive to the $U$ Stokes parameter.  The full RF band was
divided into 3 sub-bands, 32--36 GHz, 29--32 GHz, and 26--29 GHz.
Each of the channels was detected by a separate microwave
correlator, labelled J1, J2, and J3, respectively. The constant
rotation of the instrument about its vertical axis allowed for
simultaneous detection of both the $Q$ and $U$ Stokes parameters.

\polar\ employed a phase-sensitive detection technique. The
relative phase of the two arms of the correlation radiometer was
modulated at approximately 1 kHz.  An analog lock-in circuit
produced the actual ``in-phase'' polarization signal. However, the
instrument simultaneously locked into the chop frequency shifted
by $\frac{\pi}{2}$, yielding pure noise. We obtained one of these
``quadrature-phase channels'' (QPC) for each of the three in-phase
channels (IPC). The quadrature-phase channels were a good monitor
of the noise in our system, and as such were employed in several
places throughout the analysis pipeline. We use the terms
J1i, J2i, and J3i to refer to the three in-phase channels, while
J1q, J2q, and J3q refer to the corresponding quadrature-phase
(noise) channels.

The data were sampled at 20 Hz and continuously recorded to files,
each containing
precisely 9000 samples (7.5 minutes) of data; these files are one
of the fundamental data units to be discussed throughout this
paper. The instrument was calibrated daily with a 3-mil thick
dielectric sheet. We did not calibrate the instrument when the
weather was poor. The calibration was accurate to about 10\%,
based upon laboratory measurements of the dielectric sheet
properties; see \cite{odell02} for details.  Each {\it section}
of data is defined as the longest period of observations between
calibrations. Approximately fifty such sections were collected
throughout the observing season, each some 2--24 hours in length.

\polars\ constant rotation rate allows us to characterize its
response to a polarized signal as follows: \beqa{polaresponse}
\y(t) \; & = \; & I_0 \: + \: C \cos{\omega t} \: + \: S\sin{\omega t} \notag \\
\; & \; & \quad \; + \: Q \cos{2\omega t} \: + \: U\sin{2\omega t} \: + \: n(t)\qquad
\eeqa
where $\omega = 2\pi f = 0.2055
\;rad\; s^{-1}$ was the (angular) rotation frequency. The constant
offset $I_0$ is due to coupling of the unpolarized total power
signal into the polarization channels via the nonzero
cross-polarization of the instrument. This offset term was
typically 10-100 mK, depending on the channel; during good weather
its stability was better than 0.6 mK per hour. $C$ and $S$ are
signals modulated at the rotation frequency (referred to hereafter
as \onephi\ signals), and can be caused by various types of ground
pickup and other systematic effects.
Via \eqn{polaresponse}, linearly polarized signals will produce a
signal in the data stream at twice our rotation frequency,
henceforth called the \twophi\ frequency.

\section{Observations and Data Selection}\label{s:cuts}
During the observing season, \polar\ observed 24 hours per day
over a two month period, yielding roughly 750 hours of data.
However, the observing season contained a large diversity of
weather conditions, and this led to a correspondingly large
diversity of data quality. Developing robust data selection
criteria was one of the most critical tasks in the data analysis
pipeline. Because of the data's diversity, we were not able to
arrive at a single selection criterion; rather, we developed a
battery of conditions that the data were required to meet before
being accepted.  Many effects can conspire to contaminate the
polarization signals, be they instrumental, atmospheric, or
celestial.  We wish to flag and remove any data with a
non-cosmological contribution to our signal that mimics a
cosmological signal in a way that we cannot account for and
remove.

\begin{table}
\caption{\label{cuttable}
Amounts of data cut at various steps in the data selection process.}
\begin{ruledtabular}
\begin{tabular}{cccc}
\vspace{-1mm}
\textbf{Cut Type} &
\textbf{Hours}&
\textbf{Hours}&
\textbf{Indiv.}\\
 &
\textbf{  Surviving}\footnotemark[1] &
\textbf{\small  Cut}\footnotemark[2] &
\textbf{\small  Hours Cut}\footnotemark[3] \\
\hline
Full Data Set & 746.5 (100\%) & 0 (0\%) & 0 (0\%) \\
Instrument & 629.3 (84.3\%) & 117.3 (15.7\%) & 117.3 (15.7\%) \\
Celestial & 270.9 (36.3\%) & 358.4 (48\%) & 434.3 (58.2\%) \\
Data-based & 152 (20.4\%) & 118.9 (15.9\%) & 478 (64\%) \\
Length-based & 78.1 (10.5\%) & 73.8 (9.9\%) &  N/A \\
\end{tabular}
\end{ruledtabular}
\footnotetext[1]{The amount of data left after that
cut and all the cuts above it have been applied.}
\footnotetext[2]{Amount of data cut at that stage.}
\footnotetext[3]{Amount of data that would
have been cut if that particular criterion were the only one applied.}
\end{table}

The rest of this section will describe the various criteria we
established in order to robustly separate out contaminated data.
There were three basic types of cuts used: instrument-based,
celestial, and data-based.  Instrument-based cuts were those in
which our systems were not functioning properly or led to
unanalyzable data. Celestial cuts were those in which some
non-cosmological celestial source was in a position to contaminate
our data, such as the sun, moon, or galaxy. Data-based cuts were
those which used noise properties of the data itself to assess
contamination. \tbl{cuttable} lists all of the cut criteria used
in the \polar\ analysis; the entries in the table will be described
in more detail throughout the rest of this section.

We flagged data at the file level; that is, we either kept or cut
individual 7.5-minute length data files.  This proved convenient
as well as ensured good noise stationarity over an individual
file, while allowing for a good estimation of the low frequency
noise, necessary for our mapmaking algorithms.  A related question
to the cutting timescale is whether or not to cut all channels
simultaneously versus separately. An important fact of our
experiment was that some of the effects contaminating the data
were frequency dependent; in particular, channel J3 (26--29 GHz)
often showed evidence of a spurious signal when the other two
channels did not. For this reason, we performed the data-based
cuts on each of the channels individually; the instrumental and
celestial cuts had no frequency dependence, and hence were applied
equally to the three channels.

\subsection{Instrument-Based Cuts}
We performed two instrument-based cuts. The first cut was to
remove any files during which the system wasn't rotating for the
entire file, or the rotation was unnaturally slow or jittery. This
cut removed 1.8\% of the data. The other issue was dew formation
on the optics, a common problem in ground-based experiments. When
the relative humidity was high for an hour or more, moisture
condensed on the vacuum window, leading to a spurious polarization
signal. The mechanism was likely that ambient radiation scattered
off the layer of water, leading to the spurious signal. Dew
contaminated necessitated removal of 14\% of the data.

\subsection{Celestial Cuts}
Because the feedhorn accepts some radiation from all angles, a
celestial source will contribute a nonzero amount of radiation
when it is anywhere above the horizon.  We are concerned with
celestial objects leaking radiation through a sidelobe in our
beam, or the main lobe if the source ever gets close enough to the
local zenith.  The sun, moon, and planets, as well as point
sources and diffuse radiation from the galaxy are all possible
contaminants.

\subsubsection{Solar System Objects}
We used the measured beam profile (see K02) to determine
attenuation of the microwave emission from the
sun, moon, and planets as a function of angle from the beam (which
equals altitude for our zenith-staring experiment).
Treating the sun as an 11,000 Kelvin
blackbody with an angular width of 0.5\deg FWHM \cite{zirin91},
and conservatively assuming 1\% polarization, we find contamination
from the sun is less than $< 0.4\, \uK$ for elevations less than 70\deg.
Similarly, the lunar model by \cite{keihm83} shows that the moon can be
treated as a blackbody of approximately 240 K in its brightest
phase, with an angular diameter of 0.5\deg\, and $\lesssim 1\%$
polarization. Again assuming $1\%$ polarization,
lunar emission falls below 0.05 \uK\ at zenith angles
greater than 40\deg.  Using cut elevations of 70\deg for the sun and 40\deg
for the moon removed a sizeable 42\% of the data.

In principle, contamination by the planets was also possible.
The largest planetary signal was estimated to be $\sim 0.01\, \uK$ in total intensity.  This was
due to Jupiter, which reached a maximum of 65\deg elevation during our observations.
Including the fraction of the source that is polarized will reduce this number
even further. Thus in the case of \polar, planetary radiation can be safely neglected.

\subsubsection{Galactic Foreground Emission}\label{galcuts}
Finally there is the question of emission from the galaxy, either
from diffuse or point sources. \polar\ had a sensitivity to point
sources of roughly 2 \uK/Jy.  Using the WOMBAT point-source
catalog for 30 GHz \cite{wombat}, we calculate that there are
approximate 40 sources that passed through our beam
contributing greater than 1 \uK\ antenna
temperature in intensity. Only three sources contribute greater than 10 \uK.
We chose not to perform a point-source based cut as detecting
these sources, while challenging, would be both interesting and
relatively easy to identify as galactic versus cosmological in
origin.

The observations described were conducted over a wide range of
galactic latitudes and therefore there is a potential for diffuse galactic
contamination, especially at low latitudes \cite{ben96,dav96}.
Galactic synchrotron emission can be up to $70\%$ polarized
\cite{rl79}. No maps of polarized synchrotron emission exist at
$30$ GHz, and extrapolation of measurements at lower frequencies,
\eg \ \cite{bs76}, is not a reliable probe of
synchrotron polarization at $30$ GHz due to Faraday rotation.
Although the unpolarized intensity is apparently not correlated
with the polarized intensity as shown in \cite{adoc02}, we
attempt to limit our susceptibility to synchrotron emission by
only using data corresponding to Galactic latitudes
$|b|>25\deg$.

\begin{figure}[tb]
\begin{center}
\renewcommand{\subfigcapmargin}{24pt}
\subfigure[]{\includegraphics[width=3in]{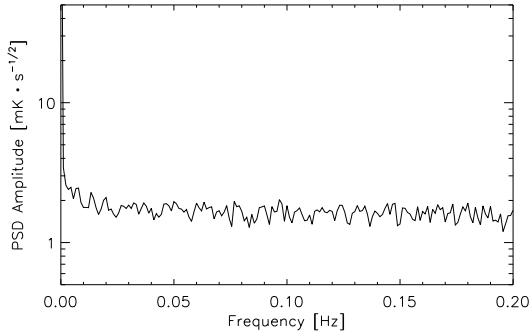}}
\subfigure[]{\includegraphics[width=3in]{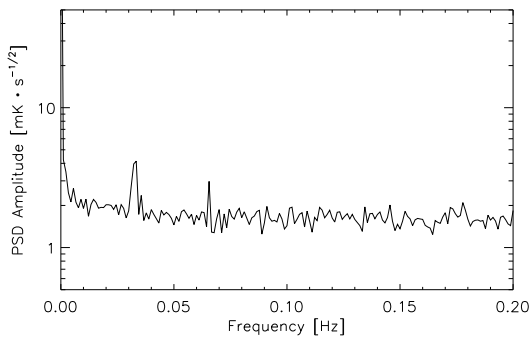}}
\caption{\label{lowfreqpsd} Low frequency power spectrum of two
sections of data for channel J2i.  Panel (a) shows a power spectrum of the time stream
when the weather is good and systematic effects are
low. Panel (b) displays a similar period when clouds appeared; there is a dramatic
appearence of spikes at both the \onephi\ and \twophi\ frequencies.  This motivated our using
the strength of the \onephi\ peak as a cut statistic. True celestial polarization signals
will appear at the \twophi\ frequency only.}
\end{center}
\end{figure}

\subsection{Data-Based Cuts}\label{deepercuts}
Together, the instrument-based and celestial cuts removed
approximately 64\% of the original 750-hour data set.  Cursory
analysis, however, immediately revealed that spurious signals
still remained in this reduced data set at
unreasonable levels. This is not surprising given that the
weather in Wisconsin is highly variable in the spring, and most
of the radiation entering the system was atmospheric emission.
Especially important was the presence of clouds, which
can mimic a polarized signal to our radiometer\citep{cothesis}.
We therefore follow the standard
practice of deriving selection statistics based upon
the data itself, to serve as a measure of the quality of observing conditions.

\subsubsection{The \onephi\ Cut}
Signals modulated only at the \twophi\ frequency correspond to true
polarization signals; thus, a signal that has response at both
\onephi\ and \twophi\ cannot correspond to a true celestial
signal.  This can be seen in the power spectrum of the
data (\fig{lowfreqpsd}).  The upper (a)
panel shows a roughly featureless power spectrum
taken during a period of good weather for
channel J2i. But as the weather worsens, due to clouds and/or
increased humidity and water vapor, features at the \onephi\ and
\twophi\ frequencies appear (\fig{lowfreqpsd}(b)).
This motivates a study of the correlation between \onephi\ and
\twophi\ signals in the time stream.

\begin{figure}[tb]
\begin{center}
\includegraphics[width=2in]{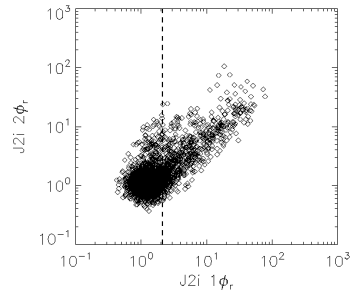}
\caption{\label{one_v_two} \twophir\ vs \onephir\ for the
data channel J2i.  Periods contaminated by sun,
moon, and dew have been removed.  The vertical dashed line shows
the $\onephir \ < \ 2.1$ cut level. }
\end{center}
\end{figure}

We computed the \onephi\ and \twophi\ values for each data file
\emph{relative to the noise floor}\footnote{We found it useful
to compare the height of these harmonics relative to the noise floor, to remove
the effect of variable radiometer noise.}, denoted as
\onephir\ and \twophir\ respectively.
\fig{one_v_two} displays these
data plotted against each other for channel J2i, with both
instrument-based and celestial cuts applied. The high degree of
correlation between the two quantities is striking;
this strongly motivated the use of the \onephir\ level as
a cut statistic.
In order to determine a logical \onephir\ cut level, we compared
our data to the behavior of white noise. Monte-Carlos showed that
this statistic is $1.0 \pm 0.27$ for white noise,
distributed roughly as a Gaussian. We opted to cut
whenever $\onephi_r > 2.1$; this is more than $4\sigma$ from the
mean for white noise. Implementing this requirement cut an additional
15\% of the data.  However, even after instituting this cut there remained
unphysically high values of \twophir\ in the data, motivating
additional cuts.

\subsubsection{The Zeta Cut}
Because the autocorrelation function
and power spectrum of any data set form a Fourier transform pair,
the information in one is the same as in
the other. For instance, a rise in 1/f noise leads directly to a
higher ``floor'' in the autocorrelation function.  Any signal in
the data will lead to a non-trivial autocorrelation function.
We used that fact to our
advantage and defined the following statistic for each data file:
\beq{e:zeta}
\zeta \ \equiv \
\frac{\sum_{lag=1}^{1000}{C(y_{in})}^2}
{\sum_{lag=1}^{1000}{C(y_{quad})}^2}
\eeq
where $y_{in}$ denotes
data from an in-phase channel and $y_{quad}$ denotes data
from the corresponding quadrature-phase channel.
Simulations showed that a reasonable astrophysical signal
(of, say, less than 100 \uK rms), would have a negligible
contribution to \Zeta\ for \polar.
\eqn{e:zeta} has an intuitive explanation; it is roughly the integral
of the power spectrum between 0.01 Hz (near our lowest observable frequency)
and 10 Hz (our Nyquist frequency), weighted by $1/f$,
so low-frequency drifts cause \Zeta\ to increase rapidly.
As such drifts are generally indicative of poor atmospheric
conditions, \Zeta\ proves to be a sensible cut statistic.

\begin{figure}[tb]
\begin{center}
\includegraphics[width=3in]{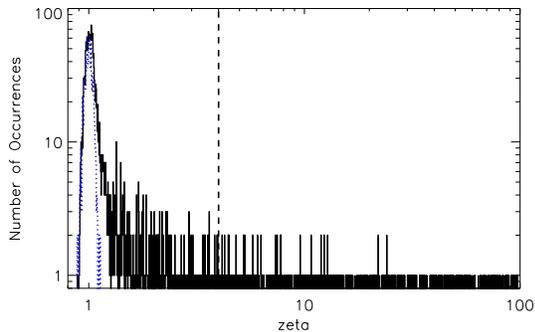}
\caption[Distribution of the \Zeta\ Variable] {\label{zeta_hist}
Distribution of the \Zeta\ Variable.  The solid (black) curve
shows the distribution of \Zeta\ for channel J2, with the basic
sun, moon, and dew cuts applied.  The dotted (blue) curve shows
the \Zeta-distribution for simulated white noise run through our
anti-aliasing filter.  The vertical dashed curve is the cut level
of \Zeta=4.0. }
\end{center}
\end{figure}

\fig{zeta_hist} shows a histogram of \Zeta\ values throughout the
season (with the basic sun, moon, and dew cuts applied).  Also
shown in the plot is a model of our data stream with no 1/f noise
or signal of any kind.  As can be seen, this distribution is
peaked around 1.0, with more than 99.9\% of the data lying
below 1.2. However, any deviation from white noise will
rapidly change the $\Zeta$ value of the data.  We found that
requiring $\Zeta < 4$ cut about 3\% of the data
over the other cuts.  Varying the actual cut
level between $\Zeta < 2$ and $\Zeta < 5$ had little effect
on the final results.

\subsubsection{Outliers in the Time-Ordered Data} Occasionally, birds, planes,
electrical noise, etc.,  would cause short-lived yet large spikes
in the data stream. For each data file and channel, we calculated
the mean and standard deviation, and recorded how
many standard deviations the first, second, and third strongest
outliers were from the mean.  We cut any data file whose
\emph{second largest outlier} was more than $5\sigma$ from the
mean of that data.  Data files with one strong outlier alone rarely
occurred, because our 5 Hz filters in conjunction with our 20 Hz
data acquisition frequency guaranteed that even a delta-function
signal should have a width of at least a few samples in the data
stream; thus the second strongest outlier proved a more robust
indicator for such spurious effects. This cut removed
$\app \ 1.2\%$ of the data in addition to all
previous cuts.

\subsection{Duration-Based Cuts}
In order to ensure that our surviving data segments were obtained during
long, contiguous periods of good weather, we instituted a battery of three
duration-based cuts.
First, we required that if a data file were to survive, both its neighbors
had to survive as well. Second, we required a minimum of eight
consecutive data files (one hour of data) to survive in order to keep any piece of
data. Lastly, we required a minimum of
three hours total to survive in a \emph{section}
\footnote{A section of data is a data segment that was taken
without interruption (this was typically about one day of data).}
of data for any data in that section to be retained;
in contrast to the cut above, this three hours
was not required to be contiguous.
This cut was instituted because of the offset removal step
of the analysis pipeline, described in \sct{offsetremoval}.
This algorithm takes place on a per-section basis,
and effectively removes all information from very short sections of
data; in the end we found it simpler just to remove these short
sections of data. Together, the duration-based cuts removed about 10\%
of the full data set.

\section{From Data to Maps}\label{s:mapmaking}
We now proceed to construct $Q$ and $U$ maps of
the post-cut data set, a process which does
not remove any cosmological information \cite{teg97a}.
Many authors have written on the mapmaking problem
and all the nuances that can arise during its solution:
Wright (1996) \cite{wright96};
Tegmark (1997), hereafter T97a \cite{teg97a}; Bond, Jaffe, Knox (1998), hereafter
BJK \cite{bjk98}; and Stompor et al. (2001), hereafter S01 \cite{stomp01}.
We use the following notational conventions. When discussing a vector,
we use a lowercase boldface letter (\eg, $\y$) to represent it.
Similarly, matrices are represented by uppercase boldface letters (\eg, $\W$).

We employ the technique of \emph{Minimum Variance Mapmaking} throughout
this work, defined briefly as follows; the interested reader is referred
to T97a for a thorough discussion of this topic.
We begin with a set of $n$ observations $\y={y_1\ldots y_n}$ of signal plus
noise, such that $\y = \A\x+\n$, where $\x$ is the underlying map, $\n$ represents noise,
and $\A$ is the pointing matrix that encodes the mapping from data space to map pixel
space (see \cite{wright96} for examples). Let $\N$ represent the data noise covariance matrix,
where $\N \equiv \langle \n \n^t \rangle$. The best estimate $\xt$
of the true map $\x$ is given by
\footnote{This estimate of $\xt$ is typically termed the
``COBE-style solution'' \cite{jansen92,wright96}.}
\begin{subequations}
\label{mmeq}
\beqa{RODy}
\label{mmeqa} \xt & = & [\A^t\N^{-1}\A]^{-1} \A^t\N^{-1} \; \y \\
\NN & = & [\A^t\N^{-1}\A]^{-1} ,
\eeqa
\end{subequations}
where $\NN$ is the \emph{map noise covariance matrix} of $\xt$.
Note that the \emph{total map covariance matrix}, $\C$, is the sum of signal
plus noise: $\C \equiv\ \Sc + \Sig$, where $\Sc \equiv \eval{\x
\x^{t}}$ is the theory covariance matrix. The
total covariance matrix will be used in \sct{s:likes}.
This method is quite general, and can be applied to any type of problem where a linear
combination of data is made in order to determine some physical
quantity. Whenever the noise of said data isn't white, this is
the best approach. For \polar, we exploit this technique no
fewer than four different times throughout the analysis pipeline.

\subsection{Constructing the File-Maps}
\label{s:polarmaps}
The data pipeline begins with binning the data into our map coordinates, via \eqn{mmeq}.
We formed a small \emph{file map} for each 7.5-minute (9000-sample) data file,
for each channel and in both $Q$ and $U$.  The ``map'' we calculated for each data
file consisted of the $Q$ and $U$ parameters for whatever map pixels were observed during that
particular file.  Because of our scan strategy, a zenith drift scan at $\delta =
43\deg$, our maps are one-dimensional in right ascension (for each of $Q$ and $U$).
We pixelized our maps with 180 pixels of width 2\deg\ in RA,
with the first pixel arbitrarily centered at RA = 0\deg.
The pointing matrix $\A$ was formed according to \eqn{polaresponse};
note that this matrix is not sparse, and is significantly more complicated in the case
of extracting both $Q$ and $U$ rather than the usual intensity $T$ (see \cite{cothesis} for
further details).

We performed binning on each data file separately because
our noise was only stationary over periods of tens of
minutes. In principle, even nonstationary noise can be treated with the
standard mapmaking formalism, but the algorithms are
unacceptably slow because the covariance matrix
$\N$ has no special properties except that it is symmetric, and the
central rate determining step of the mapmaking algorithm of
\eqn{mmeq} is in calculating $\N^{-1}$.
For stationary noise, however, the corresponding noise covariance
matrix has the special property that it is both symmetric and
\emph{Toeplitz}; that is, each upper left--lower right diagonal is
the same.  Toeplitz matrices can be inverted in \order{n_t^2}
time, as opposed to \order{n_t^3} for general $n_t \times n_t$
matrices \cite{golub96}. However, the memory requirements are
still \order{n_t^2} because the inverse of a symmetric Toeplitz
matrix is not in general Toeplitz. For \polar, $n_t=9000$, so
simply holding the inverse matrix takes $\sim 324$ MB of memory
for double-precision arithmetic.  While not prohibitive, this
memory requirement motivates further study of approximation
techniques.

We use a slight modification of the circulant matrix approximation
introduced in \cite{teg97c}, which notes that a typical noise
covariance matrix $\N$ is nearly \emph{circulant}.  In this case,
$\N = \N_c + \N_s$, where $\N_c$ is the circulant component of
$\N$ and $\N_s$ is the non-circulant component. As $\N_c$ is the
major contributor to $\N$, \cite{teg97c} lets $ \N \rightarrow \N_c$ in
\eqn{mmeqa}; the equation for the covariance matrix becomes
significantly more complicated. This solution is valid in that it
still produces an unbiased, nearly-optimal approximation of the
map, at the expense of slightly increased noise. It has the
advantage that circulant matrices can be inverted in \order{n_t
\ln{n_t}} time.

However, it is possible to enact the circulant matrix approximation in
a more streamlined way.  Using the definitions that
\begin{subequations}
\label{RODeff} \beqa{yeff}
\yeff \ \equiv \  \N_c^{-1/2} \y \\
\Aeff \ \equiv \ \N_c^{-1/2} \A \eeqa \ ,
\end{subequations}
the circulant matrix approximation becomes
\begin{subequations}
\label{mmeff}
\begin{eqnarray}
\xt & = & [\Aeff^t \Aeff]^{-1} \Aeff^t \yeff \\
\NN & \cong & [\Aeff^t \Aeff]^{-1} \ .
\end{eqnarray}
\end{subequations}
The second equation above is not precisely exact, but for our data it was correct to better than 1\%.
We state the mapmaking algorithm in this fashion because there is a very fast way to calculate
$\yeff$ and $\Aeff$. Because a circulant matrix becomes
diagonal in the Fourier domain, it can be shown that
\beq{fft_trick}
\vect{b}_{pw} \; \equiv \; \N_c^{-1/2} \vect{b} \; = \;
\ft^{-1}\left\{\frac{\fft{\vect{b}}}{\sqrt{S_N(f)}}\right\}
\eeq
where $\vect{b}$ is any vector of length $n_t$, $S_N$ is the power
spectral density that characterizes the covariance matrix $\N$,
and ``$\ft$'' denotes the Fourier Transform.
This approach is advantageous in that it requires only a simple fit to the noise
power spectrum for the time stream of interest; it is unnecessary
to explicitly calculate the resulting covariance matrix $\N$. The
resulting vector $\vect{b}_{pw}$ is simply the prewhitened
version of $\vect{b}$. We fitted our noise to a model containing
1/f noise alone:
\beq{model_PSD}
S_N(f) = \sigma^2
\left(1+\frac{f_{knee}}{f}\right) \ ,
\eeq
where $f_{knee}$ is the knee frequency of the 1/f noise.
It proved unnecessary to include the anti-aliasing filter in the model,
as it has virtually no effect on the power spectrum at the \twophi\ frequency,
where the $Q$ and $U$ data live.
We used \eqn{fft_trick} to then calculate $\yeff$ and $\Aeff$
\footnote{\eqn{fft_trick} was applied to each to each row of $\A$ to form the corresponding
row of $\Aeff$.} for all data files passing the cuts, and applied \eqn{mmeff}
to generate roughly one thousand 1--2 pixel maps in each of $Q$ and $U$, along with the
corresponding covariance matrices.

\begin{figure}[tb]
\begin{center}
\includegraphics[width=3.3in]{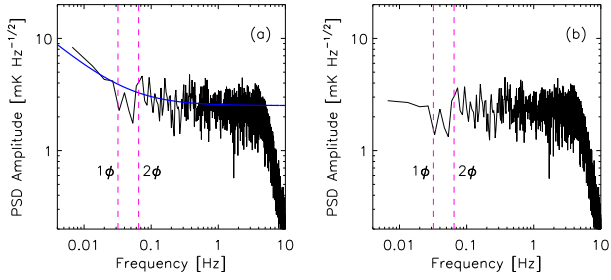}
\caption{\label{psdfilter}  Effect of prewhitening on a
sample noise power spectrum for channel J2i.
Panel (a): PSD amplitude for a sample data file, with a
knee frequency of \app\ 0.07 Hz.
The thick (blue) curve is the logarithmic-weighted fit to
the PSD. The vertical dashed lines show the \onephi\ and \twophi\
rotational frequencies.  Panel (b): Same as (a), but for the
$\yeff$ version of the data.  The effect was to whiten the
power spectrum of the data; it is equivalent to dividing the
PSD by the fitted curve.}
\end{center}
\end{figure}

\subsection{The problem of the $Q$ and $U$ Offsets}

\begin{figure}[t]
\begin{center}
\includegraphics[width=3.3in]{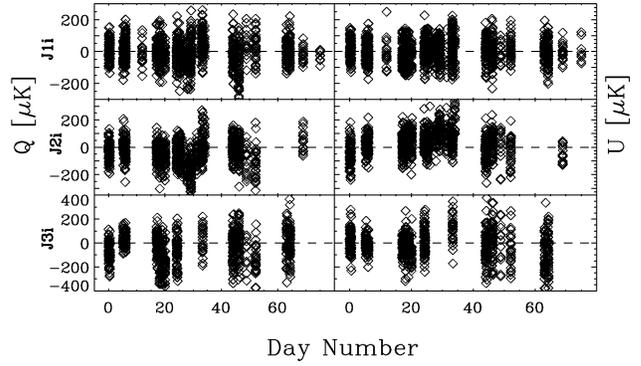}
\caption{\label{QUtimestream} The mean values of $Q$ and $U$ for
the \polar\ in-phase data versus time, after the cuts have been
applied. Each data point represents one data file. Non-Gaussian
behavior is immediately apparent at the 50-100 \uK\ level.
This behavior was not seen the quadrature-phase data.}
\end{center}
\end{figure}

At this point in the analysis, a serious issue revealed itself.
The $Q$ and $U$ data were not in general consistent with zero; there were
slowly-varying offsets in $Q$ and $U$
that varied over the course of the season, and differed between
the channels.  This fact is illustrated in \fig{QUtimestream},
which shows the derived time stream of $Q$ and $U$ values
for all the surviving data. Notice first that the data lie in
``chunks'' along the time axis; each such chunk corresponds with
a \emph{section} as defined in \sct{s:instr}.  The distribution of $Q$'s
and $U$'s for each section is consistent with a Gaussian distribution,
but those distributions are in general not centered around zero.
Notice also that there is no clear relationship between the $Q$ or
$U$ offset among the three channels.
This is in contrast to the QPC data, which were
offset-free. Thus the offset was not electronic in nature,
and also was not consistent with a celestial signal when plotted in sky coordinates.

The underlying cause of the offset was never discovered, but there
are several reasonable possibilities.  K02 hypothesizes
that the quadrupolar shape of the outer groundscreen may be to
blame \cite{kea02}. \cite{cothesis} examines the
characteristics of the offsets in great detail, and
providing several theories for the possible cause.
If we had some model of the effects that could make
successful predictions about its level, we could use it to safely
subtract out the offsets without strongly affecting our signal
recovery. However, without such a model we proceed along a different tack.

\subsection{Submap Generation}\label{s:addmaps}
It is a simple task to combine all the file maps for a single channel and Stokes parameter,
within a single section into a \emph{submap} (each corresponding roughly
to one night of good data).
Given a set of file maps to be combined, we first expand them to the full 180 pixel
map by assigning unmeasured pixels infinite noise (and hence zero weight).
The correlations between Stokes $Q$ and $U$ were negligible both within a pixel
and between pixels, and were therefore neglected.  This allowed $Q$ and $U$ to be treated
completely independently throughout the analysis.
We use the standard map combination prescription to combine the file maps into submaps
(see \eg, \textcite{qmap98c}).
Given a set of $m$ independent maps $\{\x_1 \ldots \x_m\}$
with corresponding noise covariance matrices $\{\N_1$ $\ldots$ $\N_m\}$,
the best estimate of the underlying map $\x_f$ and its
associated covariance matrix $\N_f$ are given by
\begin{subequations}
\label{addmaps}
\begin{eqnarray}
\N_f \ = \ \left[ \sum_{i=1}^m \N_i^{-1} \right]^{-1} \\
\x_f \ = \ \N_f \ \left[ \sum_{i=1}^m \N_i^{-1} \ \x_i \right] \ .
\end{eqnarray}
\end{subequations}
In this way, we combined all file maps into submaps for each
section and channel. At this step each submap was
still left with a substantial offset in $Q$ and $U$, as discussed
previously.\\

\subsection{Marginalization and Submap Combination}\label{offsetremoval}
We next employed the technique of
\emph{marginalization} to each such submap \cite{bjk98}, in order
to remove sensitivity to the mean of $Q$ and $U$ in each.  We
performed the algorithm separately for each channel.  Note
that this technique provides identical results to those obtained
using the method of \emph{virtual pixels} for mode removal,
described for instance in \cite{teg97c}.
Detailed recipes for applying the marginalization technique are
described in S01 \cite{stomp01}; we give a cursory review here
as to how it was specifically applied for POLAR. Given a submap
$\y$ (containing $Q$ or $U$ data) with covariance matrix $\N$, the
covariance matrix is transformed to
\beq{constrmatrices}
\N' \ = \ \N + \sigma_c^2 \ \Z \Z^t
\eeq
where the matrix $\Z$ contains a normalized column vector corresponding
to each mode to be removed, and $\sigma_c$ is some large number
to ensure the modes get zero weight.  For marginalization over the mean
we take $\Z = \eO$, where $\eO$ is the column vector of all 1's.
Taking the limit as $\sigma_c^2 \ \rightarrow \ \infty$, the inverse of $\N'$
still exists and is given by
\beq{Ninv_marg}
\N'^{-1} \ = \N^{-1} - (\N^{-1} \Z)[\Z^t \ \N^{-1} \Z]^{-1} (\N^{-1} \ \Z)^t
\eeq

We then applied \eqn{addmaps} to combine all the submaps into
a single map of $Q$ and $U$ for each channel, for both the IPC and QPC data.
This recipe only requires knowing the inverse covariance matrix corresponding
to each submap, given explicitly by \eqn{Ninv_marg}.
However, we still have to perform the final inversion of
\be
\Sig^{-1} \equiv \sum_{i=1}^m {\N'}_i^{-1}
\ee
to find the final noise covariance matrix $\Sig$, and $\Sig^{-1}$
is singular due to our marginalization over the means of all the submaps.
Following S01, the the final covariance matrix is taken to be
\beq{final_cov}
\Sig \ = \ (\Sig^{-1} +
\epsilon \Z \Z^t)^{-1} -\epsilon^{-1} \Z \Z^t
\eeq
where $\epsilon$ is any small positive number. In practice, it is best
to choose $\epsilon$ to be of the same order as the nonzero
eigenvalues of $\Sig^{-1}$.

The final maps for each of our three frequencies were
constructed according to the recipe given above and
presented in K01.
Qualitatively, there is no visual evidence of a common signal
among the three sub-bands, in either $Q$ or $U$. The $\chi^2$
values from each map are also not consistent with a statistically
significant signal.  Let us additionally consider the
corresponding maps made from the QPC (pure noise) channels.  We do not expect
these to contain signals, but they serve as a useful
litmus test when viewing the IPC maps: if the IPC maps differ strongly
from the QPC maps, that is evidence of either signal or some type
of contamination in the IPC maps.
However, that is not the case; none of the QPC maps
contain strong outliers, and all exhibit $\chi^2$ values
consistent with no signal.

\subsubsection{Combining the Channel Maps}\label{s:finalmaps}
Based on the fact that there is no apparent correlation between the maps
of each channel, we simplify the analysis by combining the
three channel maps into one combined map (and covariance matrix) for each
of $Q$ and $U$.
Since the CMB signal is independent of frequency (when expressed
in thermodynamic units), we are free to do this.
We would like to employ the standard map co-addition algorithm of
\eqn{addmaps} to perform this task.  However, that algorithm
assumes that the measurements made of each individual map are
\emph{independent}; if there is a systematic effect that
introduces correlations between measurements from different
channels, then we have less information than we think we do,
and this fact must be taken into account in constructing the summed map.

We first evaluated the correlation coefficients
among channels J1i, J2i, and J3i in the time-ordered data.
As expected, the correlations in the time stream were on the order
of 1\%, and were due to the very slight overlap of the three channels in frequency
space (see K02 for details).
However, it is not the time stream correlations that we so
much care about, it is the correlations between $Q$ or $U$ for the
channels.  For instance, if there were a 10\% correlation between
J1i-$Q$ and J2i-$Q$, it could be hidden in the smaller time stream
correlations.  We must therefore evaluate these correlations
directly.

\begin{table}
\caption{\label{t:iccc} Inter-Channel Cross-Correlation
Coefficients}
\begin{ruledtabular}
\begin{tabular}{ccccc}
 & $\mathbf{\expec{QQ}}$ & $\mathbf{\expec{UU}}$ &
$\mathbf{\expec{QU}}$ & $\mathbf{\expec{UQ}}$ \\ \hline
$\mathbf{\expec{J1 \ J2}_{IPC}}$ & 0.144 $\pm$ 0.034 & 0.134 &
-0.024 & 0.021 \\ $\mathbf{\expec{J1 \ J2}_{QPC}}$ & 0.005
$\pm$ 0.034 & 0.062 & -0.011 & -0.002 \\ $\mathbf{\expec{J1
\ J3}_{IPC}}$ & 0.074 $\pm$ 0.041 & 0.063 & -0.069 & -0.003 \\
$\mathbf{\expec{J1 \ J3}_{QPC}}$ & 0.023 $\pm$ 0.041 &
0.048 & -0.009 & -0.042 \\ $\mathbf{\expec{J2 \ J3}_{IPC}}$
& 0.104 $\pm$ 0.041 & 0.093 & -0.024 & -0.029 \\
$\mathbf{\expec{J2 \ J3}_{QPC}}$ & 0.066 $\pm$ 0.041 & 0.001 &
-0.054 & -0.047 \\
\end{tabular}
\end{ruledtabular}
\footnotetext[1]{The errors are the same
within each row, and we assume that the underlying distribution of
correlation coefficients is Gaussian.}
\end{table}

In order to measure these correlations, we used the $Q$ and $U$ data set
and determined the Pearson's correlation coefficient in the same way as
for the time-ordered data, but because there is so much less data,
we evaluated only one correlation coefficient for each surviving section.  We
calculated means and errors of these coefficients
by averaging from the distribution of these values, shown in
\tbl{t:iccc}. The numbers in this table are very suggestive.  For
instance, $\expec{Q \ Q}$ for all IPC channels is about the same
as $\expec{U \ U}$, suggesting a common source.  All QPC
correlation coefficients are consistent with zero, as are all
correlations of the $\expec{Q \ U}$ variety \footnote{ Except
perhaps $\expec{Q_1 \ U_3}$, but because all the other
coefficients of this type are consistent with zero, we assume it
is an outlier.}. As $Q$ and $U$ show no correlation
between them (either within a channel or between channels), we can
continue treating $Q$ and $U$ as completely independent measurements.
This is not too surprising, considering they are essentially the
$\sin{\twophi}$ and $\cos{\twophi}$ projections from each
rotation, which are orthogonal functions.  However, the
correlations between IPC channels (for the same Stokes parameter)
are \app\ 10\%, so we cannot ignore them in constructing a final
map.

Let us now combine the maps of $Q$ (or $U$) from the three
non-independent channels, armed with the knowledge of their mutual
correlations. The algorithm of \sct{s:addmaps} did not deal with
adding non-independent maps together, but it is relatively easy to
expand the methods to do so.  We treat $Q$ and $U$ separately, as
they are completely uncorrelated. Let us consider our situation
for $Q$ ($U$ will follow an identical format).

\begin{figure*}
\begin{center}
\subfigure[]{\label{finalmap_in}
\includegraphics[width=3.2in]{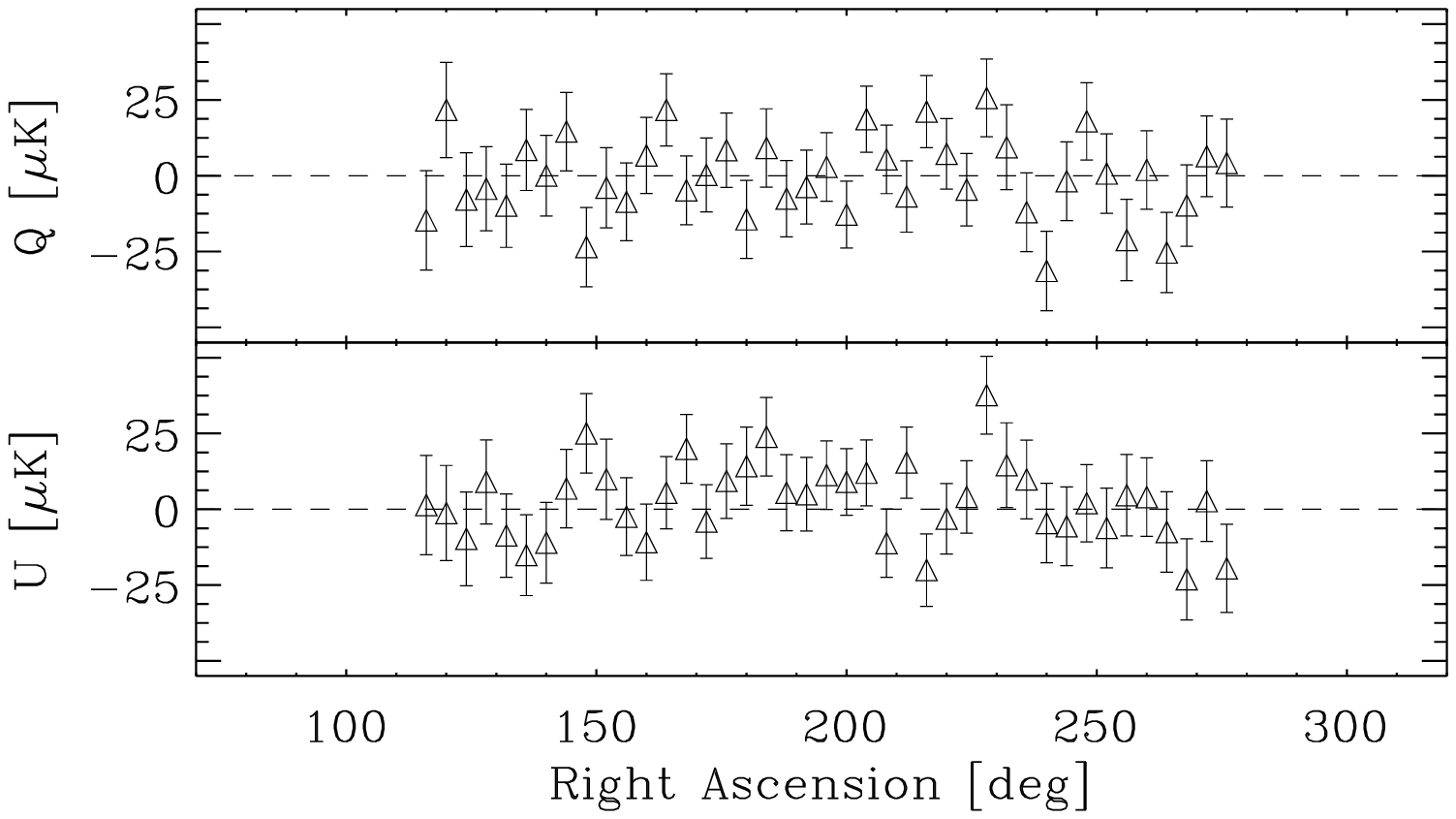}
}
\subfigure[]{\label{finalmap_quad}
\includegraphics[width=3.2in]{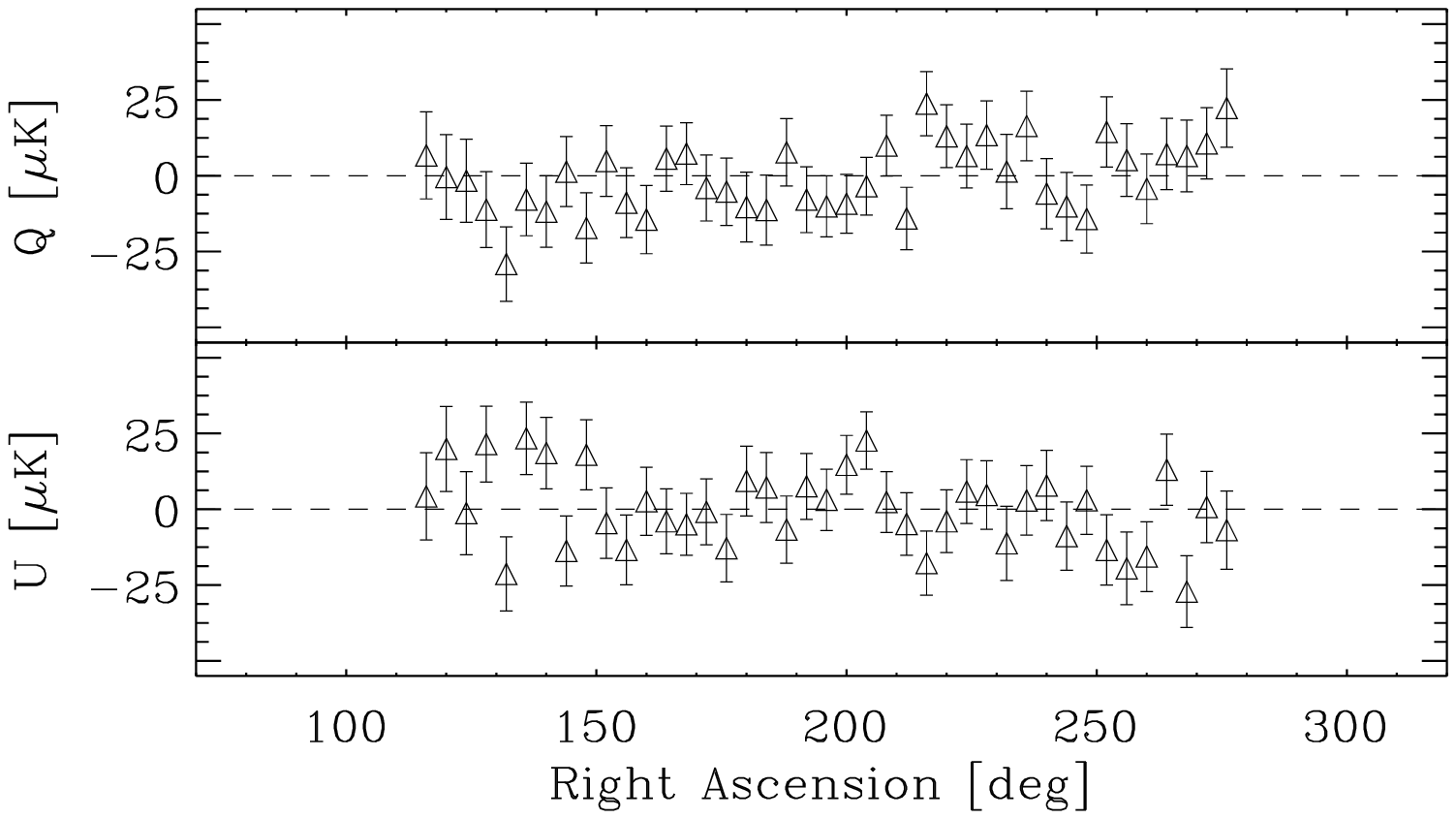}
}
\caption{\label{finalmap} Final joint-channel sky maps.  The
joint IPC maps for $Q$ and $U$ are displayed in (a), while (b)
shows the QPC maps.}
\end{center}
\end{figure*}

We have three channel maps, call them $\x_1$, $\x_2$, and $\x_3$, with their
corresponding covariance matrices $\Sig_1$, $\Sig_2$, and
$\Sig_3$. Let the correlation coefficient between $\x_i$ and
$\x_j$ be $\rho_{ij}$. We appeal to \mm\ to form the best possible
map. First, we form the concatenated map and corresponding
covariance matrix:
\begin{subequations}
\label{megamap}
\begin{eqnarray}
\x_{cat} & = & \{\x_1, \x_2, \x_3\} \\
\Sig_{cat} & = & \left[
\begin{array}{ccc}
\Sig_1 & \rho_{12}\Sig_{12} \ & \rho_{13}\Sig_{13} \\
\rho_{12}\Sig_{12} \ & \Sig_2 & \rho_{23}\Sig_{23} \\
\rho_{13}\Sig_{13} \ & \rho_{23}\Sig_{23} \ & \Sig_3
\end{array} \right]
\end{eqnarray}
\end{subequations}
where $\Sig_{ij} \equiv \sqrt{\Sig_i}\sqrt{\Sig_j}$.
We can take the square roots of the $\Sig$-matrices since they are
all positive definite, as long as we add a large offset to each
matrix (corresponding to the uncertainty in the offset, which is
formally infinite)\footnote{ For a symmetric, positive-definite,
$n\times n$ matrix $\M$, its square root is given by $\PP^t
\D^{1/2} \PP$, where $\PP$  is an $n \times n$ matrix such that
the $i^{\text{th}}$ row of $\PP$ contains the $i^{\text{th}}$
eigenvector of $\M$, and $\D^{1/2}$ is a diagonal matrix with the
square-roots of the eigenvalues of $\M$ along its diagonal. The
eigenvectors must be normalized, such that $\PP \PP^t = \I$.}.
The final full covariance matrix, $\Sig_f$, will then also have a
large offset, but again this corresponds to our marginalization
over the means of all the component submaps and is will
not affect our final results.

We now apply the \mm\ formalism to the concatenated map (which
in this case is our data vector \y\ from \eqn{mmeq}).
The pointing matrix is given by
\beq{jointpointingmatrix}
\A_{cat} = \left[ \begin{array}{c} \I_n \\ \I_n \\ \I_n
\end{array}\right] \
\eeq
where $\I_n$ is the $n\times n$ identity matrix, and $n$ is the
number of pixels in our maps.
This matrix points our three individual maps to the same final map.
Explicitly, the final joint map and covariance matrix are given by:
\begin{subequations}
\label{addchannelmaps}
\begin{eqnarray}
\x_f & = & \Sig_f \
\A_{cat}^t \Sig_{cat}^{-1} \x_{cat} \\
\Sig_f & = & [\A_{cat}^t \ \Sig_{cat}^{-1} \A_{cat}]^{-1} \ \text{.}
\end{eqnarray}
\end{subequations}
Because of the large offset each covariance matrix possesses, the
final map $\x_f$ may have some random offset to it, but it is
meaningless, and can be safely subtracted out.  Notice that in the case of vanishing
inter-map correlations, we correctly reproduce the algorithm of \eqn{addmaps}.

The final joint-maps for the IPC and QPC are shown in
\fig{finalmap}.  There is no obvious evidence of an underlying sky
signal.  In order to determine robustly if they show evidence of a signal,
we employ the standard techniques of maximum likelihood analysis and quadratic estimators,
described fully in \sct{s:likes}.

\subsection{Testing the Mapmaking Pipeline}
We next simulated the data stream in order to test our map reconstruction
pipeline. There were three primary steps involved in the
simulation process: build the underlying map, let \polar\
``observe'' this fake sky and generate data based on these
observations, and construct maps from the resulting data
set. We built the underlying sky maps out of simple sine and
cosine modes, for both total intensity ($I$) and Stokes $Q$ and
$U$. We assumed a basic flat band-power model with \app\ 10 \uK\
per band, and convolved these signals with our 7\deg\ (FWHM) Gaussian beam.
This signal level is of course highly unrealistic for true CMB
polarization; we used it merely so we could reconstruct the actual
map with good signal-to-noise (instead of merely providing an
upper limit).

\begin{figure}[tb]
\begin{center}
\includegraphics[width=3.3in]{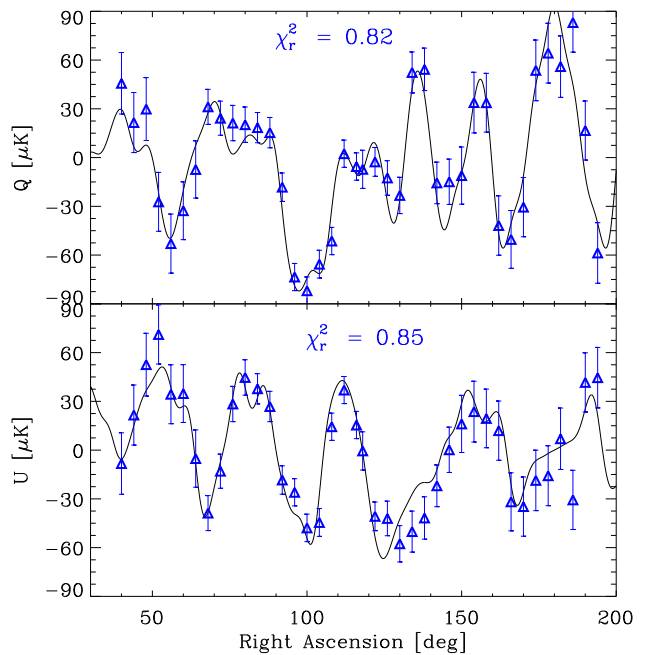}
\caption[Maps of Simulated Data] {\label{simmap} Derived
joint-channel maps from simulated data for one example
``fake-sky'' map.  The thin (black) curve is the
underlying sky map convolved with a 7\deg\ beam.  The (blue) data
points represent the derived joint-channel map, after the 31 hours
of simulated data in 5 submaps.
The good agreement illustrates the robustness of our mapmaking
algorithms and of the marginalization technique
used to remove sensitivity to means of $Q$ and $U$ from each contributing
submap.}
\end{center}
\end{figure}

To shorten processing time, we made the simulated instrument
about a factor of two more sensitive than the real \polar.
We included all IPC and QPC channels in the analysis,
but no total power channels.  We
assumed the noise was almost white, with a small amount of 1/f
noise in each channel in agreement with the true data set. We
convolved each data stream with our 5 Hz anti-aliasing filter to
ensure a resulting power spectrum that was similar to the actual
data. We also added random offsets in $I$, $Q$, and $U$ for
each section and channel, of levels consistent with those
experienced by \polar.  The results of these simulations showed
that our map reconstruction algorithms worked as expected, even
when including large random offsets for each section in the $Q$
and $U$ data, and illustrates the amazing utility of the marginalization
technique.  The $Q$ and $U$ reconstruction of one such
``fake sky'' is shown in \fig{simmap}.

\section{Constraining CMB Polarization Models}\label{s:likes}

\subsection{Limits on E and B in a Flat Band-Power Model}\label{s:fbp}
\subsubsection{Primary Likelihood Analysis}
We now address the question of the level of rms polarization in the
joint-channel maps using a standard likelihood analysis.
For the beginner, an excellent discussion of how to carry out
such an analysis is provided in \cite{bunnthesis}.

As \polar\ measured both Stokes $Q$ and $U$ simultaneously,
we are able to set limits on both $E$- and $B$-type polarization
independently. As described previously in K01, we
constrain a flat CMB power spectrum characterized by two
polarization temperatures $T_E$ and $T_B$ such that
$\l(\l+1) \Cl^X/2\pi=T^2_X$ where $X \in \{E,B\}$.
We then solve for the likelihood function $\L(\a)$, given by
\beq{like}
\L(\a) \ \propto \ \frac{e^{-\frac{1}{2} \x^t\C^{-1}(\a)\x}}{|\C(\a)|^{1/2}} \ ,
\eeq
where $\a$ is our parametrization vector $\a\equiv\{T_E,T_B\}$,
$\x = \{\q, \u\}$ is the concatenation of
the joint channel maps of $Q$ and $U$ given in the previous section,
and $\C(\a)$ is the full covariance matrix given by
the sum of the data (or noise) and theory covariance matrices:
\beq{fullcov}
\C(\a) \ = \ \Sc(\a) \ + \Sig \ .
\eeq
The construction of the theory covariance matrix $\Sc$ is rather complicated
\cite{kks,matias98,teg01a}, and has been previously presented
for POLAR \cite{kea01,cothesis}.
The data covariance matrix $\Sig$ is formed from
the covariance matrices for the joint-channel maps of $Q$ and $U$ such that
\begin{eqnarray}\label{datacov}
\Sig \ = \ \left[\begin{array}{cc} \Sig_Q & 0 \\ 0 & \Sig_U \end{array} \right] .
\end{eqnarray}
Note that $\Sig$ has a very large offset (and corresponding eigenvalue)
due to our original marginalization over the submap means,
but its specific magnitude will not affect results of the likelihood analysis.

\begin{figure}[tb]
\begin{center}
\includegraphics[width=3.5in]{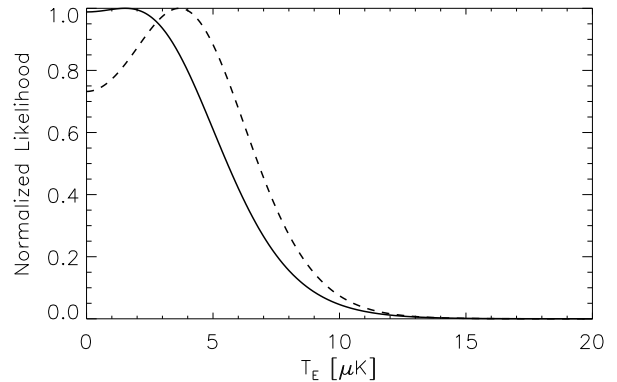}
\caption{\label{eonly} Normalized likelihood plot of $T_E$, with
the prior constraint that $T_B = 0$. The solid line is the result
for the in-phase channels, and the dashed line is for the
quad-phase (null) channels. The resulting upper limit is $T_E < \
7.7 \, \uK$ at 95\% confidence.}
\end{center}
\end{figure}

We calculated the likelihood function for all individual and joint
channel maps, both for the in-phase and quadrature-phase channels.
As reported in K01, all cases were consistent with
pure upper limits.  The combined-channel data yield 95\% confidence
limits of 10.0 \uK\ on both $T_E$ and $T_B$ (or 5.0 \uK\
stated using the modern convention for defining $Q$ and $U$).
As the $B$-polarization
at large angular scales is assumed to be so much weaker than
$E$-polarization (even taking into account lensing), we can can set
$T_B$ to be zero. It yields a 95\% confidence limit of $T_E < \
7.7 \uK$, as shown in \fig{eonly}.

\begin{figure}[tb]
\begin{center}
\includegraphics[width=3.3in]{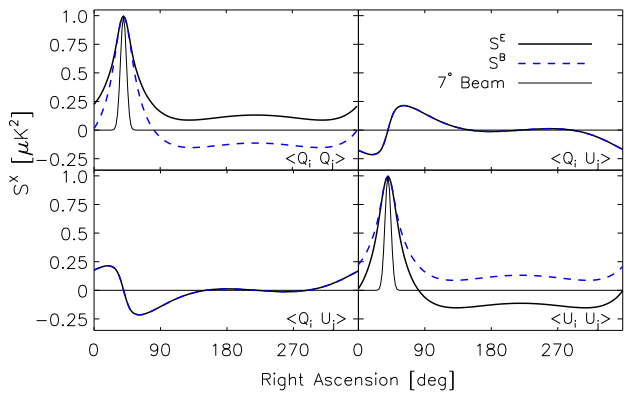}
\caption{\label{eb180} Plots of the RA = 40\deg\ row of the fundamental
signal covariance matrices $\Sc^E$ and $\Sc^B$.  The four panels
in the plot correspond to the four quadrants of the matrices as
labelled.  The thick solid line represents $\Sc^E$, the thick
dashed line represents $\Sc^B$, and the thin solid line
corresponds to a 7\deg\ Gaussian beam for reference.}
\end{center}
\end{figure}

\subsubsection{Alternative Likelihood Analyses}
To verify the analysis, we repeated it
in two different ways.  First, we performed the likelihood analysis
\emph{without} the offset removal.  The corresponding likelihood
analysis yields a spurious detection for the in-phase channels at
$(T_E,T_B) = (18,1) \pm (8,16) \, \uK$ (95\% conf.),
while the quadrature channels yield a 95\% confidence
upper limit of $7 \, \uK$ for both $T_E$ and $T_B$.
We see that the offset removal degraded the upper
limit by $\sim$ 30\%.
This is primarily due to the large width ($\sim 25\deg$)
in right ascension of the central lobe of the theory covariance matrix,
as shown in \fig{eb180},
coupled with our marginalization over the mean of each submap.
This is consistent with the fact that each submap averaged roughly 90\deg\
of coverage in right ascension.  This emphasizes to the
experimenter that having long, clean sections of data is key to
obtaining the best final noise possible, especially if any
kind of mode removal may take place in the analysis.

As a further consistency check, we performed the entire analysis in such a way that
the three channels were co-added directly in the time-ordered data.
This technique automatically takes into account any correlations
that may be present between the channels.  One simply co-adds the
time streams with their inverse noise weightings in order to obtain
a time stream with the minimum possible noise.  Offset marginalization
was still done on the submaps, but since we had already combined
all the channels, a single offset was removed for $Q$ and $U$
for each section, rather than one for each channel.
When the corresponding likelihood
contours were calculated, the upper limits remained, but were
degraded to about 12 \uK.  This makes sense considering we
subtracted only two offsets per section, as compared
with six in the primary analysis.
Because the offsets were not perfectly correlated among
the three channels, there was residual power left over in the maps
due to the imperfect co-addition of the channel offsets; it was
this phenomenon that led to the slightly degraded upper limits.  This
result notwithstanding, this analysis is noteworthy in that it
shows our inter-channel correlations were not a severe problem.

\subsection{Limits on Polarization Power Spectra}\label{qe}
CMB anisotropy and polarization are typically characterized by the
temperature fields $T$, $Q$, and $U$. These quantities can be
combined to form six measurable power spectra: $TT$, $EE$, $BB$,
$TE$, $TB$, and $EB$ (see for instance \cite{teg01a}). Our
analysis of these spectra is presented in \cite{adoc02}; we give a
brief summary of it here.  Note that, in contrast to the rest of this paper,
temperatures in this section are reported using the modern convention of $Q
= (T_x-T_y)/2$, as discussed in \sct{s:conv}.

\begin{table}
{\centerline{Table III: POLAR-DMR Power Spectrum$^{(a)}$}}
\medskip
\centerline{\label{atable}%
\begin{tabular}{lrrr}
\hline \hline
\multicolumn{1}{l}{}                      &
\multicolumn{1}{c}{$\l_{\rm eff}\pm\delta\l$}    &
\multicolumn{1}{c}{$\delta T ^2\pm \sigma\> [\mu K^2]$}&
\multicolumn{1}{r}{$\delta T\> [\mu K]^{(b)}$}         \\
\hline
$T$         &15.6$\pm$6.6     & 487.0$\pm$270.6     &22.1$^{+7.4}_{-5.5}$\\
$E$         &12.6$\pm$4.5     & -4.9$\pm$ 16.0     &$<$~3.3(5.2)\\
$B$         &12.6$\pm$4.5     &  6.9$\pm$ 16.0     &$<$~4.8(6.3)\\
$X$         &14.0$\pm$4.8     & -18.4$\pm$ 34.3     &$<$~4.0(7.1)\\
$Y$         &14.0$\pm$4.8     &  -0.1$\pm$ 34.3    &$<$~5.9(8.3)\\
$Z$         &11.4$\pm$2.9     &-25.0$\pm$ 15.8     &$<$~(~2.6)\\
\hline \hline
\end{tabular}
}
\smallskip
\noindent{\small $^{(a)}$All values reported using the convention
$Q \equiv (T_x-T_y)/2$, as discussed in \sct{s:conv}. Table
reproduced from \cite{adoc02}.}\\
\noindent{\small $^{(b)}$Values in parentheses are 2-$\sigma$ upper limits.} \\
\end{table}

As \polar\ had very weak sensitivity to temperature anisotropy
data, we used the COBE-DMR data for the 53 GHz and 90 GHz bands
averaged together as discussed in \cite{adoc02}. We analyzed the
spectra in five bands of width $\ell=5$, but subsequently averaged
them into a single band to increase the signal-to-noise ratio.  We
employed the method of quadratic estimators to evaluate the
band-powers (see \eg \cite{bjk00,teg01a}), using the concordance
model ($\Omega_{Lambda} \sim 0.7$, $\Omega_{mat} \sim 0.3$, $h \sim 0.7$)
as input for the theory covariance matrices \cite{concordance}.
All band-powers were
consistent with upper limits, the results of which are shown
in Table 6.  Note that the
2$\sigma$ upper limits on $E$ and $B$-type polarization
obtained with quadratic estimators
are 5.2 and 6.3 \uK\, respectively, in good agreement
with the 2$\sigma$ upper limits of 5.0 \uK\ obtained from
the maximum likelihood analysis.

This technique has the added benefit of explicitly calculating the
band-power window functions, while for a maximum likelihood
analysis, they are less straightforward to evaluate
\cite{knox99}.  However, there is a new twist on band-powers when
it comes to polarization;  there is always some amount of leakage
into the desired power spectrum from the other five power spectra.
In principle, it is possible to choose priors for the quadratic
estimators such that fourteen of the fifteen leakages are zero,
but the much discussed $E$-$B$ leakage remains \cite{teg01a},
although it can be kept to a minimum at the price of a modest
increase in error bars.  This leakage is a direct result of
``ambiguous'' modes existing in the scan strategy; the more
ambiguous modes, the worse will be the leakage \cite{bunn02}. Sky
coverage that is large, two-dimensional and well-connected will
have few ambiguous modes; but \polars\ one-dimensional scan strategy
leads to virtually \emph{all} of the $E$ and $B$ modes being
ambiguous.

\begin{figure}[tb]
\begin{center}
\includegraphics[width=3in]{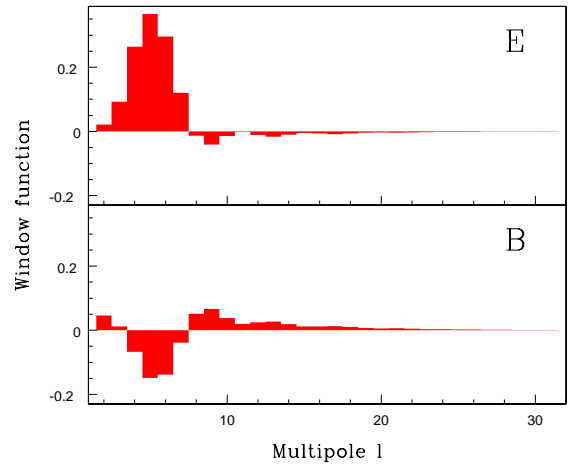}
\caption{\label{Ewindow} The window function for $E$ used to
estimate the polarization band-power near $\l = 5$, shown as a
representative example. The $B$ window functions are exactly the same as
for $E$, but with $E$ and $B$ switched.  There is significant leakage
of $B$-power into the $E$-estimate because of our one-dimensional
scan strategy. In general, a scan done over a two-dimensional
region with a large sky coverage will have a much better $E$-$B$
separation and narrower window functions. The width of the window
function scales with the inverse of the sky patch size in its
narrowest dimension.}
\end{center}
\end{figure}

\fig{Ewindow} shows the $E$ window function for a single $\l$-band
for $ 2 < \l < 8$ for \polar\ as an example; the window function
was generated using the minimum leakage techniques discussed
above. The leakage of $B$ into $E$ is exactly symmetric for $E$
into $B$, thus we show only the $E$ window function. It is clear
that there is a very large leakage from $B$ to $E$ and vice versa.
Therefore, the upper limits for $E$ and $B$ cannot be taken
separately; we constrain merely the average of the two power
spectra.

An early reionization
leads to a peak in both the $EE$ auto-correlation and $TE$
cross-correlation spectra at large angular scales
\cite{matias97,bk98}; this effect was observed recently
in the $TE$ power spectrum by WMAP \cite{kogut03}.  Unfortunately,
our $EE$ and $TE$ upper limits are still significantly higher than
any possible reionization peak.  For $\tau=0.4$, a typical concordance
model gives $\delta T_{EE} \sim 1$ \uK\ and $\delta T_{TE} \sim 3$
\uK, in comparison with our limits
\footnote{Stated using the convention $Q=(T_x-T_y)/2$ as discussed in \sct{s:conv}.}
of 5.2 \uK\ on $\delta T_{EE}$ and 7.1 \uK\ on $\delta T_{TE}$.
In comparison, WMAP recently reported $\tau = 0.17 \pm 0.04$ (68\% conf) \cite{kogut03}.

\subsection{Constraining Polarized Synchrotron Emission}
Astrophysical synchrotron emission is generally polarized at a
level from 10--50\% \cite{bs76,dav96}. Our non-detection of any
signal implies that we can place constraints on polarized
synchrotron at 30 GHz, but this is not as straightforward as it
may seem.  As discussed in \sct{galcuts}, we cannot simply
extrapolate from low frequency polarization measurements such as Brouw
and Spoelstra \cite{bs76} because of Faraday depolarization.
Extrapolation of the unpolarized Haslam 408 MHz data
\cite{haslam82} to our frequencies using a spectral index of -3.1,
assuming a polarization fraction of 10\%, and smoothing with a
$7\deg$ beam yields an rms of about 10 \uK
\footnote{\label{QUcon}Stated using the convention $Q=T_x-T_y$ as discussed in \sct{s:conv}.}.
Thus, our data favor both a steep spectral index as suggested by recent
studies \cite{platania99} and a polarization fraction of less than
10\% at large angular scales.  This issue is explored more fully in \cite{adoc02}.

\section{Discussion}\label{s:discussion}
To summarize our results thus far, using the \polar\ data we have placed limits
on CMB polarization power spectra in the context of both a flat
band-power model and a concordance model, and we have discussed
the lack of substantial synchrotron contamination in our scan
region.  We showed in \sct{s:fbp} that the offset
removal technique only degraded our limits by
$\sim 30\%$, which is not so bad considering that without it, no
limits would have been possible.

If we had integrated
for a much longer time and/or our detectors had been much more
sensitive, how well could we have actually done in determining $E$
and $B$ on large angular scales given our scan strategy?
We attempted to answer this
question via an additional simulation. We used CMBFast
\cite{cmbfast} to generate power spectra for a concordance model
universe with no $B$-modes (\ie, only scalar perturbations) and an
optical depth to reionization of $\tau=0.2$; ideally we should
recover only $E$ modes from the analysis.  We used the HEALPix
package \cite{healpix} to generate twenty sky realizations of $Q$
and $U$, and observed each with a nearly perfect \polar, one that had
a sensitivity of 0.01 \uK\ for each of 180 2\deg\ wide pixels in
our $\delta=43\deg$ RA strip. We then performed our likelihood
analysis on these map realizations limited to our scan region, and
found that we recovered roughly equal amounts of $E$ and $B$ power.
This is consistent with the window functions generated in our
quadratic estimator analysis of \sct{qe}.  On a positive note, we
did recover the correct amount of total $E$ and $B$ power;
that is, the recovered $C_\l^E + C_\l^B$ equaled
the initial $C_\l^E$ placed in the maps.

Ultimately, \polar\ was limited both by its instrument sensitivity
and atmospheric limitations to integration time.  The atmosphere
was also most likely responsible for the time-varying polarization offset.
Finally, our one-dimensional scan strategy prevented
discrimination between E-modes and B-modes. Future experiments
that effectively deal with these issues may well be able to glean
the CMB polarization signals on large angular scales, and hence
uncover a wealth of new information with which to increase our
understanding of inflation and the details of the Big Bang.

\begin{acknowledgments}
We are grateful to Josh Gundersen, Lloyd Knox, Ed
Wollack, Matias Zaldarriaga, Slade Klawikowski, and Phil Farese
for many useful conversations. BK and CO were supported by NASA
GSRP Fellowships.  BK acknowledges support from the NSF
Astronomy and Astrophysics Postdoctoral Fellowship program.
\polars\ HEMT amplifiers were graciously
provided by John Carlstrom. This work has been supported by NSF
grants AST 93-18727, AST 98-02851, and AST 00-71213, and NASA
grant NAG5-9194.
\end{acknowledgments}

\end{document}